\documentclass[twocolumn,showpacs,preprintnumbers,amsmath,amssymb,apsrev]{revtex4}

\usepackage{graphicx}
\usepackage{dcolumn}
\usepackage{bm}

\begin{document}

\preprint{}

\title{Inelastic impact of a sphere on a massive plane:
nonmonotonic velocity-dependence of the restitution coefficient}

\author{Hunter King}
\author{Ryan White}%
\author{Iva Maxwell}%
\altaffiliation[Also at ]{Physics Department, Mt. Holyoke College, S. Hadley, MA.}
\author{Narayanan Menon}
 \email{menon@physics.umass.edu}
\affiliation{
Department of Physics, University of Massachusetts, Amherst, MA 01003 U.S.A.}

\begin{abstract}
We have studied the coefficient of restitution, $\eta$, in normal collisions 
of a non-rotating sphere on a massive plate for a range of material parameters, 
impact velocity and sphere size.  The measured coefficient of restitution does not 
 monotonically vary with velocity.  This effect is due to dynamics that 
occur during the finite duration of impact: the contact time varies as a 
function of velocity is comparable 
to the time-scales of the vibrational modes of the plate.  The measured effect is robust and is expected 
to be ubiquitous in fluidized granular media. We 
also find that $\eta$ is a decreasing function of particle size, a dependence 
that is not captured by existing models of impact.
\end{abstract} 


\maketitle

Collisions of macroscopic objects - such as a ball with the floor - are typically 
inelastic: some fraction of their total translational kinetic energy is
siphoned off into viscoelastic work, plastic deformations, vibrations 
of the objects, and into producing sound. After careful experiments
on normal collisions of spheres, Newton \cite{Newton} suggested that
the degree of inelasticity could be characterized by the ratio $\eta =- v\prime/v$,
where $v$ and $v\prime$ are the relative velocities before and after impact. 
The ratio $\eta$, called the coefficient of restitution, was at first thought 
to be a constant whose value was determined solely by the geometry and the 
material properties of the colliding objects. It is now well-known that $\eta$ 
also depends on the relative velocity of impact: experiments as well as theoretical 
models \cite{Johnson, Goldsmith} indicate that  $\eta \rightarrow 1$ as $v \rightarrow 0$ i.e. the 
gentler the impact, the closer it is to an elastic collision.   In this 
article we study a particularly simple inelastic collision, that of a 
sphere colliding with a massive wall, and present data that show for 
the first time that $\eta$ is nonmonotonic in $v$ and is a decreasing 
function of the radius, $R$, of the sphere.  The data suggest that computing
 $\eta(v)$ requires a fuller consideration of the dynamics of the colliding 
objects in the finite duration of the impact.

The starting point for most models of normal inelastic collisions is the Hertz 
solution to the static problem of a sphere that is being 
pushed into a wall \cite{LandauElasticity}.  This solution -- which specifies the stress field 
in terms of compression of the sphere, its radius $R$ and the elastic moduli 
of sphere and wall -- is also assumed to obtain at any instant 
during a collision, under the condition that the velocity of impact $v$ 
is much less than the speed of sound in the solids.  To compute the 
coefficient of restitution, a model for what is judged to be the dominant 
dissipation mechanism supplements the Hertzian specification of the 
elastic force.  In recent calculations \cite{KuwabaraKono,Brilliantov, 
Oppenheim}, the dissipation mechanism has been modeled by a viscous damping term 
that is linear in the local strain rate. This yields the prediction 
$\eta _{visc}(v) \sim 1-CR^{-1}v^{1/5}$ where $C$ is a material-dependent
constant.  A different calculation 
that attributes the dissipation to plastic deformation \cite{Johnson, 
Thornton} predicts $\eta _{plastic}(v)\simeq 1.18(v/v_{y})^{-1/4}$ for $v>v_{y}$, 
the velocity at which the yield stress is first exceeded. 

Experiments on ball-ball and ball-plane collisions \cite{Goldsmith}
generally show that $\eta$ decreases with $v$, in qualitative agreement 
with the theoretical expectation. At high impact velocities the data 
are limited in range but moderately good agreement has been claimed 
\cite{Johnson} with a $v^{-1/4}$ dependence.  At lower impact 
velocities the situation is less clear: the data of Labous et al. \cite{Rosato}
for collisions between nylon beads are not fit very well by either $\eta _{plastic}$
or $\eta _{visc}$. The data of Hatzes et al. \cite{BridgesMNRAS}
on collisions of smooth ice spheres with ice bricks have been fit by $Cexp(-\gamma v)$.
Falcon et al. \cite{Falcon} find $\eta$ almost independent of $v$ for collisions of a carbide 
sphere with a steel surface.  
They point out that better agreement with their data is obtained 
with a dissipation model that is sublinear in strain-rate. Furthermore, 
the models of $\eta(v)$ yield different dependencies on size 
of the impinging sphere with $1-\eta _{visc} \propto 1/R$ 
whereas $\eta _{plastic}\propto R^{0}$.  A recent review of 
simulational schemes \cite{Schafer} for granular materials 
catalogues several simulation models in which $\eta$ increases, 
decreases, or is independent of $R$. Experiments
 that winnow down this wide range of choices are 
currently lacking.  Thus, it is our view that inspite of some 
high-quality experiments, the available data pool do not 
allow a decisive experimental validation of models
 for the size or velocity dependence of $\eta$. 
 
We have studied normal collisions of a non-rotating 
sphere on a massive wall while trying to explore a range of 
material parameters, impact velocities and sphere size. Here we 
report results for collisions on two surfaces, 
the first is a surface-ground steel disc $2.5 cm$ in thickness 
and $22 cm$ in diameter resting on a $1.25 cm$ thick sorbothane 
pad to damp vibrations in the plate, the second is a granite 
optical table $120\times 80\times 30cm$ in size.  The lateral extent of 
these surfaces is large enough to eliminate end-effects in 
the impact \cite{Sondergaard}. We have used steel, brass, aluminium, copper
and plastic (delrin) spheres of radius $R=0.47$cm. The brass spheres 
were varied in size from $R=0.3125$ to $0.9375$cm, a factor of 27 in mass. 
\begin{figure}[t]
\includegraphics [width=.5\textwidth]{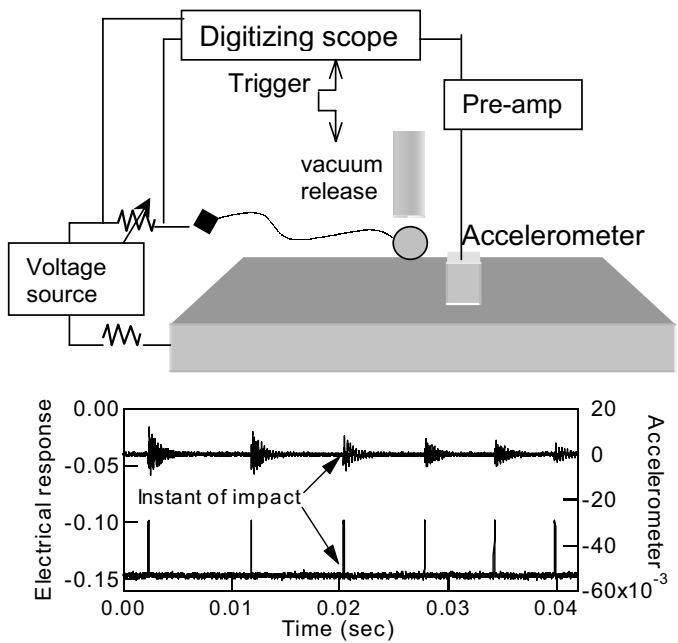}
\caption{\label{setup}(A) Schematic of setup, not to scale. (B) Timing traces 
for a sequence of 6 bounces of a brass ball on a steel plate.  
The upper trace is from the accelerometer and the lower one 
from the electrical circuit.  Both measurements are delayed 
with respect to the vacuum release to acquire a short sequence 
of bounces at a digitization rate sufficient to get $\sim 10$ 
points per contact time, $t_{c}$.}
\end{figure}
The ball is held by a vacuum and dropped without 
spin from a height of about $1 cm$ onto a massive 
plane surface \cite{Louge} by using a solenoid valve to release
 the vacuum (see Fig.1). The ball bounces repeatedly on the plane and 
finally comes to a stop.  The voltage pulse that releases the 
vacuum also triggers acquisition of the times of 
successive impacts. An
accelerometer mounted to the plane about $5 cm$ away from the 
location of the impact detects elastic waves excited by the 
impact. The instant of the $i^{th}$ impact, $t_{i}$, is obtained from the 
leading edge of the accelerometer pulse (to within a fixed 
time-lag of $ \approx 10 \mu s$).  When both sphere and plate are metallic, 
then a second determination of $t_{i}$ is 
obtained by applying a small dc voltage between ball and 
plate and finding the instant when the circuit closes. The 
electrical method is more sensitive and allows us to measure 
slower impacts; it also yields the duration of the contact, $t_{c}$.  
We have directly verified that electrostatic forces are 
negligible in the collision since our results are unchanged 
when the applied voltage is varied by a factor of 120, or 
when an ac voltage is used. Where both measurements are 
possible they yield the same result, as seen in Fig. 1.  
Given a set of collision times ${t_{i}}$ , $\eta$ at the ith bounce can be determined as 
$\eta(v_{i})=  -v_{i+1}/v_{i}= \frac {t_{i+1}-t_{i}}{t_{i}-t_{i-1}}$. 
\begin{figure}
\includegraphics[width=.5\textwidth]{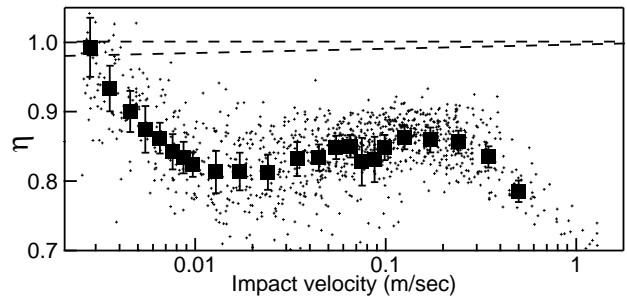}
\caption{\label{raw data} $\eta$ versus impact velocity,  
$v(m/s)$, plotted on a log-scale, for a brass sphere with $R= 0.47 cm$ 
bouncing against a steel plate.  The small dots represent 
1130 individual collisions, taken over ~100 launches of a sphere.  
The solid squares are averages of these data taken in 
logarithmically-spaced bins. The error bars are the standard 
deviation about these values whereas the dashed lines show 
the precision of individual measurements.}
\end{figure}
In Figure 2 we show the variation of $\eta$ with impact velocity 
for a brass sphere on steel.  The cloud of small 
points represents raw data while the solid 
squares are averages of these data.  
The novel and striking observation is that the velocity 
dependence of $\eta$ is non-monotonic; there is a 
range of velocities in which the collisions become more 
elastic as the impact becomes harder.   This is at 
variance with the theories described above which prescribe 
a monotonic increase towards the limit of  $\eta(v=0)=1$.  
It might appear surprising that this non-monotonic 
dependence has not previously been observed, 
since our collision geometry is fairly typical and the 
precision of some previous measurements (e.g. Ref. \onlinecite{Falcon})
is comparable to ours.  Our understanding 
of this apparent inconsistency is that resolving a broad and 
shallow features seen in Fig. 2 
requires a much larger number of data points than were 
taken in previous measurements, and that these data be 
gathered over a large range in $log v$.  The scatter 
in the raw data of Fig. 2 is not dictated by our precision 
in measuring $\eta$ (indicated by the dashed bars in the figure),
 but by bounce-to-bounce variability due to slight 
 imperfections or asperities falling within the area of 
 contact of these macroscopically smooth objects: only 
 averaging over repeated bounces reveals the underlying 
 behaviour.
 
\begin{figure}[t]
\includegraphics[width=.5\textwidth]{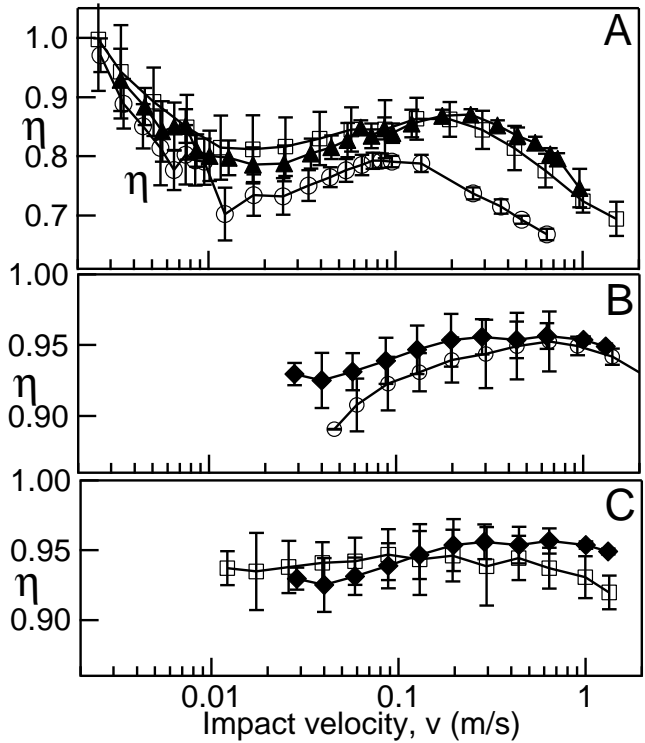}
\caption{\label{eta comparison} $\eta$ versus log $v$ (m/s) 
for (A) brass (squares), aluminium ($\blacktriangle$), and copper ($\circ$) spheres bouncing on a steel plate; (B) delrin on 
granite ($\blacklozenge$) and steel ($\circ$); and (C) brass 
(squares) and delrin ($\blacklozenge$) bouncing on granite. $R= 0.47 cm$
 in all these data.}
\end{figure}
In Fig 3 we show that the non-monotonic 
behaviour is extremely robust and can be seen for impacts 
between several pairs of materials.  Fig 3A shows $\eta$ as a 
function of velocity for several metals on a steel plate.  
While the magnitude of the inelasticity and the position of 
the minimum and peak in the data vary from one material to 
another, the overall trends in $\eta(v)$ are maintained.  
In Fig 3B we show $\eta$ for collision of a plastic (delrin) 
sphere on steel and granite surfaces.   Once again, $\eta(v)$ clearly 
shows a peak, though due to the fact that we are not able to use 
our electrical method of detection, we are unable to go to very small $v$. 
The fact that the data are displaced from each other demonstrates 
the important role of the plate, even when there is a considerable 
mismatch in the elastic modulus of the materials (delrin 
being much softer than either steel or granite).  Finally, in 
Fig. 3C, we compare the impact of brass and plastic spheres
on the granite surface.  We emphasize that the granite plate 
is much thicker than typical containing walls used in experiments 
on granular media. 

In all the cases above, the lowest impact velocity is determined either 
by the precision of our technique, or by the ultimate
 contact of the ball with the plate.  (Ref. \onlinecite {Falcon} argues that the 
last stages of this process are an elastic oscillation of the 
ball and plate under gravity).   The highest velocity we use is determined 
by the impact speeds at which we first start to observe tiny plastic 
indentations of ball or plate.   When we remain below this velocity 
there are no  \emph{visible} plastic deformations, however, we 
frequently change spheres and surface-grind the steel plate to guard against
ageing.  (This does 
not guarantee that microscopic plastic events do not occur in the 
collision.)  We have tried to eliminate adhesion between sphere and 
plate by repeatedly cleaning both surfaces between launches of the ball.  
Our results are unchanged when the experiments are done in a dry $N_{2}$ 
atmosphere and when the sphere and plane are both held at high 
temperature to expel adsorbed water and volatiles.   The viscous 
drag of the air is also negligible: in the extreme case of 
a plastic sphere at $v=100cm/s$, weight/stokes drag $\approx 2\times 10^{4}$.
Thus we believe we are close to an 
experimental idealization of the impact problem in which the important 
forces operative are gravity and the elastic forces of the media.
Why then is there such a discrepancy between theory and our observations?

\begin{figure}[b]
\includegraphics[width=.5\textwidth]{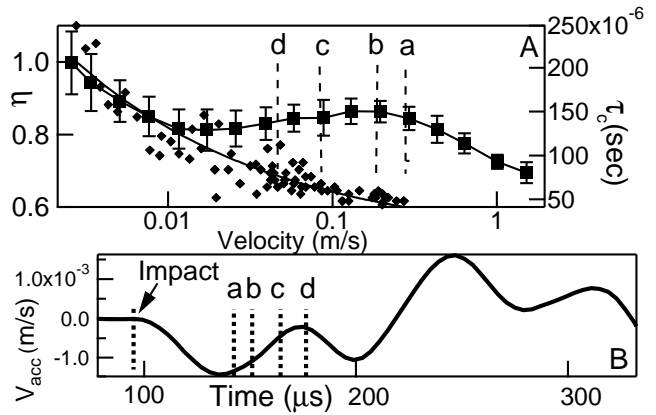}
\caption{\label{timescales}(A)  $\eta$ ($\blacksquare$, left axis) and time of 
contact, $t_{c}$ ($\blacklozenge$, right axis) versus impact velocity, $v$.  
The dashed lines labeled a, b, c, d refer to four individual collisions with 
impact velocity spaced about the peak in $\eta$. (B) Vertical velocity of 
the plate versus time following the impact.  (We show only one of the four traces since they are identical except 
for amplitude)  The instant the ball leaves the plate - as determined by loss 
of electrical contact - is labeled for each of a,b,c,d.  The plate is clearly 
in a different phase of oscillation for collisions below, above, and at the 
peak of $\eta(v)$.}
\end{figure}
We believe that the answer lies in dynamical effects that occur 
during the collision.  In Fig. 4A we show measurements of the contact 
time, $t_{c}$, of the sphere with the plane which varies with the 
impact velocity as $t_{c} \sim v^{-1/5}$, in 
approximate agreement with the contact time predicted from an 
elastic, Hertzian collision (as has recently been seen \cite{Quere} in liquid droplets, 
where deformations are large and non-Hertzian).  In Fig. 4B, we show the velocity 
of the plate as a function of time with $t_{c}$ marked on the 
time-axis for various impact velocities.  It is evident that as $t_{c}(v)$ 
changes, the phase of motion of the plate at the instant the ball leaves 
the plate can change substantially.  Thus the peak observed in $\eta(v)$ 
could conceivably be viewed as an elastic mode of the plate slinging 
the ball upward. The minimum, likewise, could correspond to the 
plate receding downward at the instant the ball leaves the plate.  
Since we are not able to measure the acceleration at the location 
of the impact we do not have a direct verification of this, but 
the data of Fig. 4 makes this explanation quite compelling. 

The effect on $\eta$ of the crossing of these two time scales -- $t_{c}$ and the 
vibrational modes of the plate -- is likely to be quite generic since the speed of 
sound in most homogenous solids does not vary by too large a factor.  Increasing 
the thickness of the plates will only shift this crossing to lower impact velocities. 
Furthermore, no matter how massive the plate, it is possible that 
surface waves on the plate, and modes of the ball, will produce similar effects. The 
idea that flexural and Rayleigh modes of the plate can play a significant role in the 
impact is not new \cite{Zener}, however, similar attention has not been paid to bulk 
modes.  Likewise, it has long been known \cite{Raman} that elastic vibrations can 
contribute to $\eta$ even without any further dissipation mechanism.  These ideas have 
been elaborated in recent continuum \cite{Zippelius} and lattice \cite{Hayakawa} 
simulations of normal impacts of discs against rigid walls.

\begin{figure}[t]
\includegraphics[width=.5\textwidth]{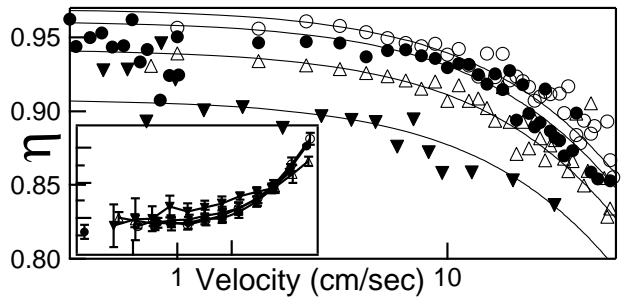}
\caption{\label{masses}$\eta$ vs. impact velocity, $v$, on a log axis, 
for the impact on a steel plate of 4 sizes of brass sphere: $
R= 0.3125 (\bigcirc), 0.47 (\bullet), 0.625 (\triangle)$, and $0.9375$ cm ($\blacktriangledown$).
The inset is a plot of $(1-\eta)/R^{1/2}$ vs. $v$, a scaling that achieves a 
reasonable collapse of the size dependence.}
\end{figure}
We have also varied the size of the sphere for brass on steel impacts.  We show in 
Fig. 5 data for $\eta$, plotted against $v$, for 4 different radii of brass spheres.  The 
data show a clear trend for $\eta$ to decrease as R increases.  In the inset to the figure 
we show that the dependence on radius is consistent with a scaling of $1-\eta \propto R^{1/2}$. 
The measured dependence is inconsistent with the size dependence of $\eta _{visc}$, 
which displays an increase in $\eta$ with increase in $R$, as well as with $\eta _{plastic}$, 
which is independent of $R$. These observations, too, appear consistent with
vibrations of the plate and sphere being important loss mechanisms, however, a
quantitative theory is clearly necessary. Labous et al. \cite{Rosato} state that their data for collisions of nylon 
spheres shows an increase in $\eta$ with size consistent with a scaling of $1-\eta \propto R^{-1/2}$, 
a trend opposite to that shown in Fig 5.   The variations in eta for their different sizes, 
however, are close to the scatter in the data so it difficult to ascertain 
whether our results are in contradiction. 

In conclusion, we have presented data that reveal an unexpectedly complex, nonmonotonic, 
functional dependence of coefficient of restitution on impact velocity. We have made 
measurements over a broad range of impact velocity, materials, and particle sizes and find this behaviour to be quite robust.  The origin of this 
nonmonotonic behaviour lies in the fact that the characteristic modes of vibration in the 
objects participating in a collision are comparable to the contact time in an impact. 
Experiments are in progress to achieve a situation where these time scales are well-separated.  
However, in general, we expect dynamical effects to be typical rather than unusual for 
collisions in granular media and to constrain the regime where quasistatic approximations 
\cite {Goldsmith, Johnson, LandauElasticity, KuwabaraKono,Brilliantov,Oppenheim,
 Thornton} will apply. Recent work \cite{Goldman, Poschel} has predicted 
macroscopic consequences of a velocity-dependent $\eta$; it remains to be seen whether 
there are new consequences that arise from the specific behaviour of $\eta(v)$ that we 
report.  It seems likely that interesting resonant phenomena might occur in sound 
propagation in granular solids that stem from the velocity dependence we find.

We thank R.A. Guyer, S. Ramanathan, L. Mahadevan and K.Z. Win for useful conversations.  
Lauren Schiller made valuable contributions to an early version of this experiment.  
We are grateful for support from ACS PRF G33730 and NSF DMR 9874833.


\begin{thebibliography}{20}

\bibitem{Newton} 
I. Newton, \emph{Principia}, see e.g. the scholium immediately following the laws of motion.

\bibitem{Goldsmith}
W. Goldsmith, \emph{Impact}, E. Arnold, 1960.

\bibitem{Johnson} 
K.L. Johnson, \emph{Contact mechanics}, Cambridge, 1985.

\bibitem{LandauElasticity} 
L. D. Landau, E. M. Lifshitz, \emph{Theory of elasticity}, Pergamon, 1970, pp 30-34 
describes the contact of two balls.

\bibitem {KuwabaraKono} 
G. Kuwabara, K. Kono, Jpn. J. Appl. Phys. {\bf 126}, 1230 (1987).

\bibitem{Brilliantov} 
N. V. Brilliantov, F. Spahn, J. M. Hertzsch, T. Poschel, Phys. Rev. E
{\bf 53}, 5382 (1996).

\bibitem {Oppenheim}
W. A. M. Morgado, I. Oppenheim, Phys. Rev. E {\bf 55}, 1940 (1997).

\bibitem{Thornton} 	
C. Thornton, J. Appl. Mech. {\bf 64}, 383 (1997).

\bibitem{Rosato} 
L. Labous, A. D. Rosato, R. N. Dave, Phys. Rev. E {\bf 56}, 5717 (1997).

\bibitem{BridgesMNRAS} 
A.P. Hatzes, F.G. Bridges, D.N.C. Lin, Mon. Not. R. astr. Soc {\bf 231}, 1091 (1988).

\bibitem{Falcon} 
E. Falcon, C. Laroche, S. Fauve, C. Coste, Eur. Phys. J. l B {\bf 3}, 45 (1998).

\bibitem{Schafer} 
J. Schafer, S. Dippel. D.E. Wolf J. Phys. I France {\bf 6}, 5 (1996) 

\bibitem{Sondergaard} 
R. Sondergaard, K. Chaney, C. E. Brennen J. Appl. Mech. {\bf 57}, 694 (1990). 

\bibitem{Louge} 
A simplified version of the setup of Louge and
coworkers, see e.g. S. F. Foerster, et al. Phys. Fluids {\bf 6}, 1108 (1994). 

\bibitem{Quere} 
D. Richard, C. Clanet, D. Quere, Nature {\bf 417}, 811 (2002)

\bibitem{Zener} 
C. Zener, Phys. Rev. {\bf 59}, 669 (1941)

\bibitem{Raman} 
C.V. Raman, Phys. Rev. {\bf 15}, 277 (1918)

\bibitem{Zippelius} 
E. Gerl, A. Zippelius, Phys. Rev. E {\bf 59}, 2361 (1999)

\bibitem{Hayakawa} 
H. Hayakawa, H. Kuninaka, Chem. Eng. Sci.
 {\bf 57}, 239 (2002).

\bibitem{Goldman} 
D. Goldman et al., Phys. Rev. E {\bf 57}, 4831 (1998).

\bibitem{Poschel} 
N. V. Brilliantov, T. Poschel, Phil. Trans. R. Soc. Lond. Ser. A {\bf 360}, 
415 (2002).

\end{thebibliography}
\end{document}